\documentclass{jhep3}
\usepackage{epsfig,cite}
\preprint{INT-PUB 04-12}
\newcommand{\be}{\begin{equation}}
\newcommand{\ee}{\end{equation}}
\newcommand{\me}{\mbox{ eV}}
\newcommand{\nue}{\nu_e}
\newcommand{\nm}{\nu_\mu}
\newcommand{\nt}{\nu_\tau}
\newcommand{\ra}{\rightarrow}
\newcommand{\s}{\sin^2 2 \theta}
\newcommand{\m}{\Delta m^2}
\newcommand{\ten}{\times 10^}
\newcommand{\su}{\sin 2 \theta}

\title{New Matter Effects in Neutrino Oscillation Experiments}

\author{Kathryn M. Zurek \\Institute for Nuclear Theory, University of Washington, Seattle, WA 98195-1550 \\Email: \email{zurkat@u.washington.edu}}
\abstract{Consequences of a new force mediated by a light scalar particle for neutrino oscillation experiments are considered.  Such a force could give rise to neutrino masses and mixings whose matter dependence is much more significant than the MSW effect.  We consider the effects of such a new force on the limits derived from oscillation experiments, and examine how the constraints on neutrino models are altered. Re-analysis of neutrino data, as well as new experiments in which large matter effects are systematically explored, is required to directly probe such physics beyond the standard model.}

\begin{document}

\section{Introduction and Motivation}

The last ten years have brought great progress with the first direct experimental evidence for the physics beyond the standard model, the discovery of neutrino masses and mixing.  In the standard model, the neutrino is a massless particle with matter interactions mediated by W and Z bosons.
Wolfenstein \cite{Wolfenstein:1978ue}, Mikheev and Smirnov
\cite{Mikheev:1985gs} were the first to show that these standard model
interactions give rise to matter dependent masses and mixings from
coherent forward scattering in the medium, and that these effects can
be significant despite the small cross-sections for such interactions.
Subsequently, many authors have considered the medium effects of
beyond the standard model neutrino interactions,with an emphasis on flavor
changing neutral interactions (for a recent analysis, see
\cite{Guzzo:2004ue,Friedland:2004pp} and references therein).  These
medium effects arise from new interactions whose energy scale is above
the standard model, and have been shown to be a small effect.
The effects of a new scalar particle coupling only to neutrinos \cite{Stephenson:1997cz}, neutrino mass arising from
matter-neutrino interactions through such a scalar\cite{Sawyer:1998ac}, and the matter
effects of a massless
string dilaton field coupled to neutrinos \cite{Horvat:1998pp} have also
been considered.  

In this paper, we wish to consider matter effects on neutrinos due to
interactions with a new, very light and weakly coupled scalar particle.  
In \cite{Fardon:2003eh}, such a new light particle was introduced as the connection between neutrinos and the dark energy, to explain the apparent coincidence that the dark energy scale $(10^{-3}\mbox{eV})^4$ is of about the same scale as the neutrino masses $\delta m_\nu \sim O(10^{-2} \mbox{eV})$.  It was also shown \cite{Kaplan:2004dq} that such new interactions with a light scalar particle could give rise to masses and mixings which are environment dependent.  Here we parameterize such a new interaction by Yukawa couplings,

\be
V = \lambda_N \phi \bar{N} N + \sum_i\lambda_{\nu_i} \phi \nu_i \nu_i + V_0(\phi),
\ee 
where $\phi$ is the new scalar field, N the nucleon field, and $\nu_i$
are neutrino fields, written here in a basis where interaction with
the scalar field is diagonal.  A light scalar field coupling to neutrinos
has been considered before, but without coupling to matter.  In a mean
field approximation, such an interaction would give rise to a neutrino
matter potential, which is an effective shift in the neutrino mass,
\be
M^{eff}_i=\frac{\lambda_{\nu_i}\lambda_N}{m_\phi^2}n_N,
\ee
where $n_N$ is the nucleon number density, $m_\phi^2\equiv V''_0(\bar{\phi})$, and $\bar{\phi}$ is the vev of $\phi$, the solution of $V'(\bar{\phi})=0$.  This minimization equation is equivalent in the mean field approximation to
\be
\lambda_N n_N+\sum_i\lambda_{\nu_i} n_{\nu_i}+V'_0(\bar{\phi})=0.
\ee
Now if $V_0$ is a sufficiently flat potential, which is the case if
$\phi$ is a light particle, then $\bar{\phi}$ will vary significantly
with $n_N$ (and $n_{\nu_i}$, which is not of interest for the
environments considered here).  This implies that there are two
sources for density dependence in $M^{eff}_i$: not only the
factor of $n_N$ in $M^{eff}_i$, but also the dependence ${m_\phi^2}(\bar{\phi}(n_N))$.  

$M^{eff}_i$ in turn modifies the effective neutrino Hamiltonian $H_{eff}$, which in a two neutrino formalism reads,
\be
H_{eff}= \frac{1}{2E}U\bordermatrix{ & &  \cr
                  & (m_1+M^{eff}_1)^2 & {M^{eff}_3}^2  \cr
                 & {M^{eff}_3}^2  & (m_2+M^{eff}_2)^2 \cr}U^{\dagger}
                +\bordermatrix{&&&\cr
                &\sqrt{2} G_f n_e &\cr
                &  &\cr},
\ee
where U is the mixing matrix, $\sqrt{2} G_f n_e$ is the usual MSW
term, $m_i$ are the vacuum neutrino masses, and $M^{eff}_i$ is some linear combination of those given in eqn.~(1.2).
The result of these new terms in $H_{eff}$ is that the effective mass
splittings and mixings (determined from the eigenvalues and
eigenvectors of the matrix), and their density dependences, are
modified.  The modification could be quite significant, as the new
 terms could yield effects potentially much larger than the standard MSW term.  Consistency with gravitational force experiments requires the coupling in the new term $\lambda_N<10^{-2} m_N/M_{Pl}$ \cite{Adelberger:2003zx}. This small coupling, however, can be compensated by $n_N$ in typical earth densities $3\mbox{g/cm}^3$, and would give rise to a potential of size
\be
M^{eff}_i= 1\mbox{ eV} \left(\frac{\lambda_\nu}{10^{-1}}\right)\left(\frac{\lambda_N}{10^{-21}}\right)\left(\frac{\rho_N}{\rho_N^o}\right)\left(\frac{10^{-6}\mbox{eV}}{m_{\phi}}\right)^2,
\ee 
where $\rho_N^o= 3\mbox{g/cm}^3$, so that $M^{eff}_i$ could potentially be much larger than the MSW term, of $O(10^{-23} \mbox{ eV})$ for typical earth densities.

In this paper we will take the point of view that these non-standard matter effects do in fact dominate over the MSW effect.  We will then consider how our interpretation of neutrino oscillation experiments shifts within the framework of these new matter effects. I wish to stress that, although such a new light particle was originally motivated in connection with the dark energy, non-standard interactions with any new light scalar particle, due to the long range nature of the force enhanced by $n_N$, could potentially give rise to new significant matter effects.  This is in contrast to the effects considered from interactions with new heavy particles, which are small due to the short range nature of the force.   

The outline of the paper is as follows.  In the first section we make model independent considerations on how the interpretation of neutrino oscillation data is changed when large non-standard interactions are considered.  In the second section we apply these new considerations to the standard three neutrino model, and discuss the implications for future experiments.  Lastly, we specifically consider the range of possible models consistent with results from LSND.

\section{New Limits from Medium Dependence}

For simplicity in the analysis, and because we wish to make our
considerations independent of a model for how $M^{eff}_i$ depends on
density, we consider here only two media: air versus rock, or more
generally, any High Density Material (HDM hereafter), which includes
any combination of rock, steel, and concrete.  We assume that the
effects of rock or any HDM on the oscillation parameters are similar,
and that they are significantly different from the air parameters.
Additionally, because the analysis here is independent of a model for
how $M^{eff}_i$ depends on density, no limits from solar neutrinos are derived.  To extract the vacuum neutrino masses and mixings from any solar neutrino data requires a model for how $H_{eff}$ and hence $m^2_{\phi}(\bar{\phi})$ depends on density; whether the masses and mixing are significantly different from the terrestrial values is highly model dependent, and hence will not be considered here.    Lastly, we assume that at the interface of two different media the sudden approximation is valid, so that the evolution of the neutrino phase is continuous.  This is valid so long as the Compton wavelength of the neutrino mass in the medium in short compared to the distance traveled in the medium. 

With these assumptions, we summarize in table~1 the experimental
signals we have to date, their HDM and air pathlengths, and limits
derived.  There are several modifications to the limits derived when
considering large, non-standard in-medium effects.  First, as
mentioned above, in order to derive limits from the solar neutrino
data, we need a model to determine how $V''_0(\bar{\phi})$, and hence
$M^{eff}_i$, depends on density.  Without such a model, we are unable to extract $\delta m^2$ and $\sin^2 2\theta$ from the solar neutrino data, as the non-standard dependences of $\delta m^2$and $\sin^2 2\theta$ on solar density are unknown.  

Second, the standard analysis of the Super-K data utilizes an up-down ratio of $\nu_e$ and $\nu_\mu$ events \cite{Toshito:2001dk}, one pathlength of which is mostly in air, and the other pathlength mostly in rock, which could potentially have very different mixings and splittings. To find the range of allowed parameters in HDM, one can instead make use of the ratio of upward stopping to upward through-going muon sample \cite{Fukuda:1999pp}, with limits as given in the final column of the table. Notice that these limits are considerably less restrictive than those of the whole set. 

Third, for negative searches partially in rock or air, $\m_{min}$ is scaled by the fraction of the path in HDM.  One can see this by noting that for small $\m L/4E$, the amplitudes are simply additive:
\be
A(\nu_x\ra\nu_y)=\langle\nu_x|U_n...U_2 U_1|\nu_y\rangle=A_n(\nu_x\ra\nu_y)+...+A_2(\nu_x\ra\nu_y)+A_1(\nu_x\ra\nu_y),
\ee
where $U_n$ is the evolution matrix in medium n, and $A_n(\nu_x\ra\nu_y)$ is the oscillation amplitude in medium n.  To determine the bounds in the medium of interest, we set all other oscillation amplitudes to zero except in the medium of interest, and we find that in medium n 
\be
P_n(\nu_x\ra\nu_y)\simeq(\m_{min} L_{n}/2E)^2,
\ee
so that $\m_{min}$ is simply scaled by the ratio of the pathlength in medium n, $L_n$, to the total neutrino pathlength.  To determine $\s_{min}$ from these null searches, we again take mixings in all other media except the one of interest to zero.  Hence, at large $\m$, 
\be 
P(\nu_x\ra\nu_y)=\frac{1}{2}\s_{min},
\ee 
so that $\s_{min}$ is unchanged.  This applies, for example, to the CHOOZ reactor experiment, where the upper bound  on $\m$ in HDM is scaled by a factor of 5-10, as the experiment is $80-90\%$ in air, but at large $\m$, $\s_{min}=0.1$ is unchanged.  Therefore, the limit on $\nue$ disappearance in HDM for small $\m$ is derived, not from CHOOZ, but from the Palo Verde reactor experiment ($\approx 95\%$ in HDM). 

Finally, we note that we have no signals, and only constraints from null searches, for the neutrino parameters in air.  Therefore, in what remains in this paper we will consider only the range of possible neutrino models admitted by the data in HDM.  We will find that the range of possible models and allowed parameters is quite different from those admitted in the standard picture with standard interactions.

\begin{table}
\hspace{-.7in}
\begin{tabular}{l|c|c|c|c|c|c|l}
  \hline
\hline
Signal & Channel & Environment & \multicolumn{2}{c|} {SI $\m_{min,max}(\me^2$)} & \multicolumn{2}{c|}{Medium $\m_{min,max}(\me^2$)} & Ref. \\ \hline
SNO & $\nue \ra \nue,\nm, \nt$ & solar-interior & $6.5\ten{-5}$& $8.2 \ten{-5}$ & Unknown & Unknown &\cite{Ahmed:2003kj}  \\
Super-K(solar) & $\nue \ra \nue,\nm$ & solar-interior & $3\ten{-5}$& $1.9 \ten{-4}$ & Unknown & Unknown & \cite{Smy:2003jf} \\
Super-K(atm) & $ \nm \ra \nu_x$ & air/HDM & $1.9 \ten{-3}$ & $ 3.0 \ten{-3}$ & $1.5 \ten{-3}$& $1.5 \ten{-2}$ &\cite{itshitsuka:2004aa,kearns:2004aa} \\
KamLAND & $\nue \ra \nu_x$ & HDM & $10^{-5}$& $7\ten{-4}$ & $10^{-5}$& $10^{-3}$ &\cite{Suzuki:2003pt}\\
K2K & $\nm \ra \nu_x$ & HDM & $10^{-3}$&no limit & $10^{-3}$&no limit& \cite{Ahn:2002up}\\  
LSND & $\nm \ra \nue$ & HDM & $4\ten{-2}$ & $ 1.2$ & $4\ten{-2}$&$1.2$ & \cite{Wolf:2001gu,Louis:pc} \\ \hline \hline
Null Search & Channel & Environment &  \multicolumn{2}{c|}{SI $\m_{min} (\me^2)$} & \multicolumn{2}{c|}{Medium $\m_{min}(\me^2)$} & Ref. \\ \hline
KARMEN & $\nm \ra \nue$ & $\sim 50\%$ air & \multicolumn{2}{c|} {$5\times10^{-2}$} & \multicolumn{2}{c|} {$0.1$} & \cite{Wolf:2001gu}\\
Bugey& $\nue \ra \nu_x $ & air & \multicolumn{2}{c|}{$10^{-2}$}&\multicolumn{2}{c|} {N/A} & \cite{Declais:1995su,Abbes:1996nc}\\
CHOOZ & $\nue \ra \nu_x $ & $\sim 80-90\%$ air & \multicolumn{2}{c|}{$7\ten{-4}$}& \multicolumn{2}{c|}{$4\ten{-3}$} & \cite{Apollonio:2002gd,Steinberg:pc} \\
Palo Verde & $\nue \ra \nu_x $ & $\sim 95\%$ HDM & \multicolumn{2}{c|}{$2\ten{-3}$}& \multicolumn{2}{c|}{$2\ten{-3}$} & \cite{Boehm:2001ik,Boehm:pc} \\
CDHS & $\nm \ra \nu_x $ & Unknown & \multicolumn{2}{c|}{$0.25$}& \multicolumn{2}{c|}{Unknown} & \cite{Dydak:1984zq} \\
NOMAD & $\nm \ra \nt $ & $\sim 60\%$ HDM & \multicolumn{2}{c|}{$0.7$}& \multicolumn{2}{c|}{$1.2$} & \cite{Astier:2001yj,Altegoer:1998gv}\\
     & $\nue \ra \nt $ & & \multicolumn{2}{c|}{$5.9$}& \multicolumn{2}{c|}{$9.8$} & \\
CHORUS & $\nm \ra \nt $ & $\sim 60\%$ HDM & \multicolumn{2}{c|}{$0.6$}& \multicolumn{2}{c|}{$1$} & \cite{Eskut:2000de,Altegoer:1998gv} \\
      & $\nue \ra \nt $ & & \multicolumn{2}{c|}{$7.1$}& \multicolumn{2}{c|}{$11.8$} & \\ \hline \hline
Future Expmt. & Channel & Environment & \multicolumn{2}{c|}{SI $\m_{min} (\me^2)$} & \multicolumn{2}{c|}{Medium $\m_{min} (\me^2)$}& Ref. \\ \hline
MiniBooNE & $\nm \ra \nue$ & HDM &\multicolumn{2}{c|}{$2\times 10^{-2}$} & \multicolumn{2}{c|}{$2\times 10^{-2}$} & \cite{McGregor:2003ds}\\
OPERA & $\nm \ra \nt$ & HDM & \multicolumn{2}{c|}{$10^{-3}$} & \multicolumn{2}{c|}{$10^{-3}$}& \cite{Weber:2001gw} \\
MINOS & $\nm \ra \nu_e,\nu_\mu,\nt$ & HDM & \multicolumn{2}{c|}{$10^{-3}$} & \multicolumn{2}{c|}{$10^{-3}$}& \cite{Weber:2001gw} \\
 \hline \hline
\end{tabular}

\caption{Range of allowed mass splittings for standard interactions (SI) versus large, medium dependent interactions. $\m_{min,max}$ defines the range of values allowed by the signal at 90\% C.L.  $\m_{min}$ is the upper bound on $\m$ at $\s=1$ at 90\% C.L.}
\label{table1}
\hspace{.7in}
\end{table}

\section{Three Neutrino Models}

Within a three neutrino framework, the experimental signals given in the table are traditionally understood as follows.  The parameter space is determined by two mass splittings and three mixing angles.  One mixing and mass splitting is determined from the solar neutrino data, which is interpreted as the MSW conversion of $\nue \rightarrow \nm,\nt$ in the sun, with large mixing and mass splitting $O(10^{-5}\me^2)$.  The other mass splitting and a second mixing angle is determined from Super-K atmospheric data, which is interpreted as the vacuum oscillation of $\nm \rightarrow \nt$, with large mixing and mass splitting $O(10^{-3} \me^2)$.  The limit on the third mixing angle comes from the CHOOZ experiment $\s_{13}<0.1$. The combination of these limits is illustrated in figure~1a.

The constraints, however, are significantly modified with the new interactions considered here in HDM.  This is illustrated in figure~1b. The limits on $\nue$ disappearance in HDM are derived from KamLAND, with the approximate allowed range $10^{-5} \me^2< \m < 10^{-3}\me^2$, which is much less restrictive the limits from SNO $6.5 \times 10^{-5} \me^2 < \m < 8.3 \times 10^{-5}\me^2$.  For $\nm$ disappearance, the Super-K upward muon data constrains the parameters in HDM, which is less restrictive in both $\m$ and $\s$ constraints than the full set utilizing an up-down ratio of events. But although we cite the upward muon data set as the best limits in HDM on $\nm$ disappearance, bear in mind that even these limits may not be a rigorous measurement of the HDM parameters, as the analysis contains contamination from a horizon bin whose pathlength is mostly air.  For a truly rigorous constraint, a reanalysis of the data would have to be conducted where the horizon bin is removed. 

\FIGURE[t]{
\centerline{a)\ \ \epsfxsize=3.5in \epsfbox{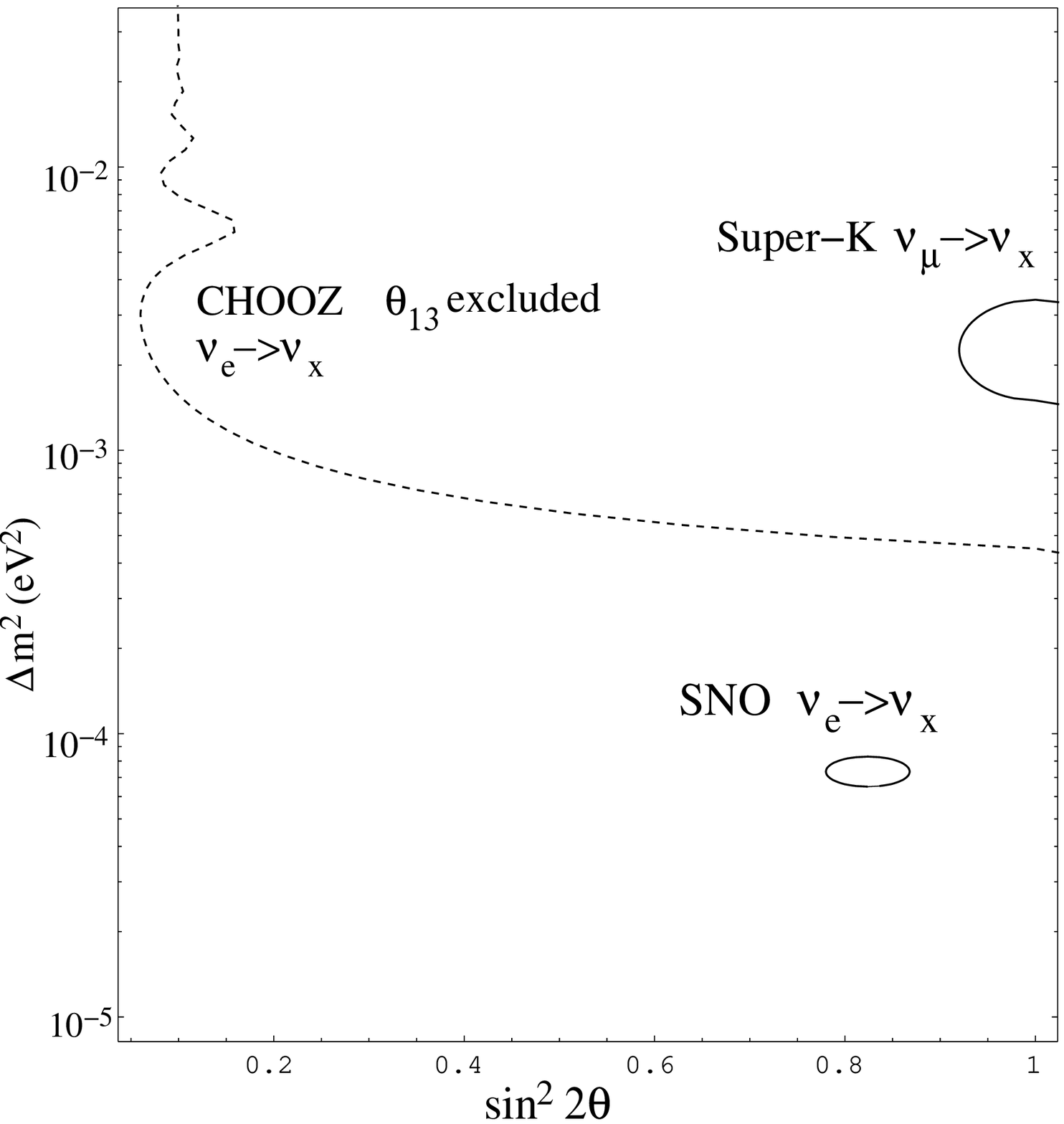} \hskip .25in
b)\epsfxsize=3.5in \epsfbox{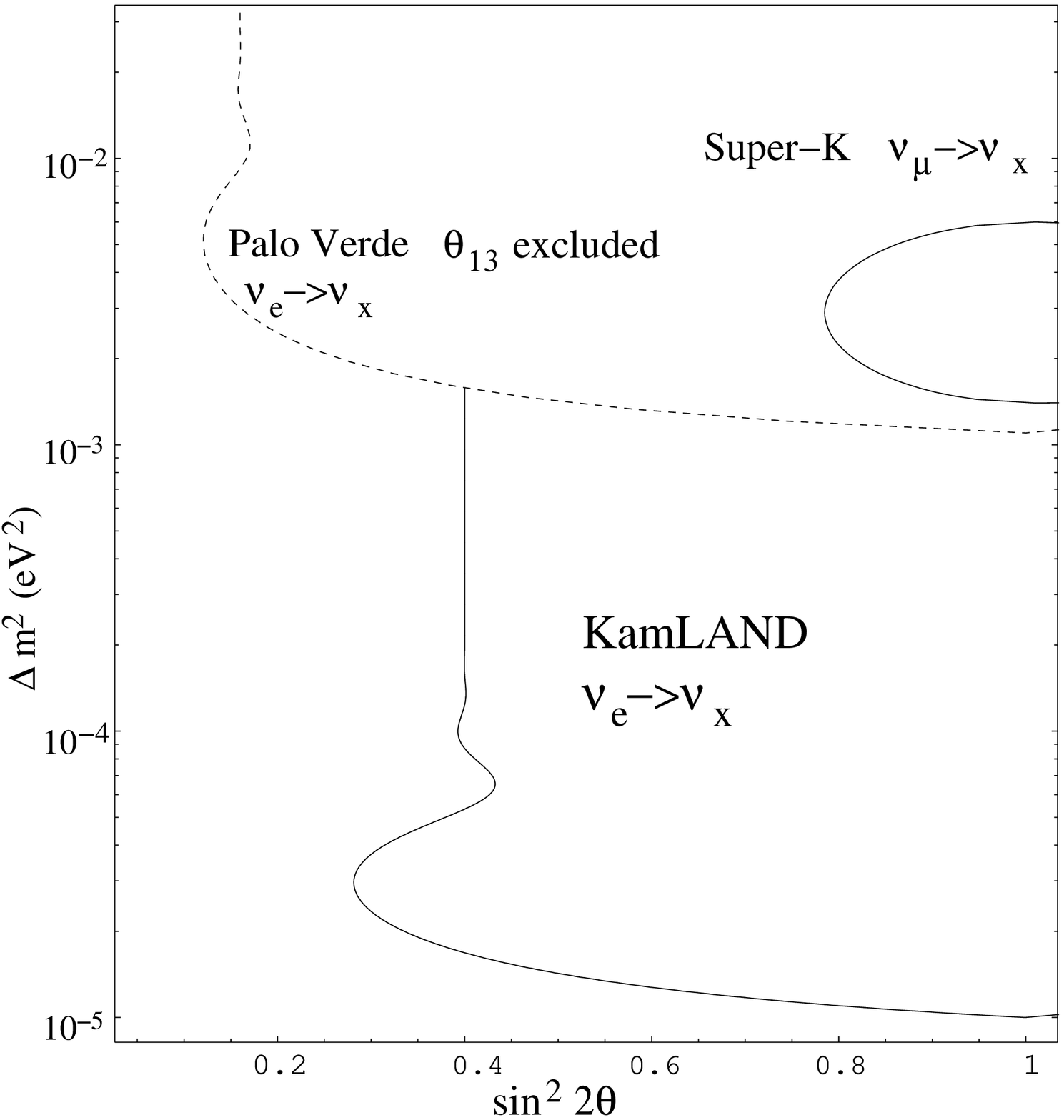} }
\caption{Allowed regions for $\nm\ra\nu_x$ and $\nue\ra\nu_x$ for a) standard interactions and b) non-standard medium dependent interactions in HDM.  The curves are parameterizations of the published allowed regions.}
}

These modifications in constraints have several implications for the interpretation of the experiments in a three neutrino model.  In a conventional analysis CHOOZ limits $\s_{13}<0.1$.  With the new interactions considered here, however, the most obvious limit on $\s_{13}$ in HDM is derived from Palo Verde.  For the allowed region of Super-K parameter space, $\m>1.5\ten{-3}$, it requires $\s_{13}\lesssim 0.5$, a rather unrestrictive limit.  In addition, the Super-K experiment has not carried out an analysis on $\nue$ disappearance mixing angles with only rock pathlengths to restrict $\s_{13}$.  The limit that we do have on atmospheric $\nue$ disappearance was derived in the full set utilizing an up-down ratio of events, with both air and rock pathlengths, in order to remove uncertainties as large as 20\% in the atmospheric neutrino flux.  For $\nue$, this-up down asymmetry $A=(U-D)/(U+D)$, where U is the number of upward-going events and D is the number of downward-going events, is measured to be quite small $A_{\nue}=-0.009\pm0.042\pm0.005$ \cite{kibayashi:2002aa}, indicating that to high certainty the observed and produced $\nue$ fluxes are the same.  

To derive, however, the allowed mixings in air and rock from $A_{\nue}$ and $A_{\nm}$ would require a new fit with six free parameters, 
\be
\m_{HDM},\ \ \s_{13}^{HDM},\ \ \s_{23}^{HDM},\ \ \m_{air},\ \ \s_{13}^{air},\ \ \s_{23}^{air},
\ee
instead of two.  As will be shown here, even an uncertainty of 5\% in the ratio of observed to produced $\nue$ fluxes for maximally oscillated neutrinos could potentially allow $\s_{13}$ to be quite large. To quantify how large $\theta_{13}$ can be consistent with Super-K's lack of observation of $\nue$ oscillation, we write the ratio of the observed $\nue$ flux to the $\nue$ flux produced in the atmosphere:

\be
\frac{\Phi_{\nue}}{\Phi^{0}_{\nue}}=\frac{\Phi^{0}_{\nm}}{\Phi^{0}_{\nue}}P(\nm\ra\nue)+\frac{\Phi^{0}_{\nue}}{\Phi^{0}_{\nue}}P(\nue\ra\nue).
\ee
We consider the limits from a maximally oscillated neutrino, which gives the tightest constraints on $\s_{13}$.  Most of such neutrinos have energy $\sim 1$ GeV and are coming directly from the opposite side of the earth ($\cos \theta \sim -1$).  For these neutrinos, the ratio of $\nm$ to $\nue$ flux is \cite{Agrawal:1996gk} 
\be
\frac{\Phi^{0}_{\nm}}{\Phi^{0}_{\nue}}\simeq2.5,
\ee 
and is quite well known (uncertainty $\sim 5\%$). Substituting this ratio into the relevant oscillation probabilities, we find
\be
\frac{\Phi_{\nue}}{\Phi^{0}_{\nue}}=1-\s_{13}(2.5\sin^2{\theta_{23}}-1)\sin^2{x},
\ee
where we are using the usual conventions $\theta_{23}$ for the $\mu-\tau$ mixing and $\theta_{13}$ for e-disappearance, and $x=\m_{atm} L/4E$.  To get a rough estimate of the limits on $\theta_{13}$ in HDM from Super-K, we note again that $\nue$ are not observed to oscillate within some fractional uncertainty $\epsilon$ in flux, so we can approximate
\be
1-\epsilon<\frac{\Phi_{\nue}}{\Phi^{0}_{\nue}}<1+\epsilon.
\ee
For neutrinos which have maximally oscillated, $\sin^2{x}=1$, we find
\be
-\epsilon<\s_{13}(2.5\sin^2{\theta_{23}}-1)<\epsilon.
\label{ne}
\ee

Likewise, we can write for $\nm$ disappearance
\be
\frac{\Phi_{\nm}}{\Phi^{0}_{\nm}}=P(\nm\ra\nm)+\frac{1}{2.5}P(\nue\ra\nm), 
\ee
so that
\be
\frac{\Phi_{\nm}}{\Phi^{0}_{\nm}}=1-(2.4\sin^2{\theta_{23}}\cos^2{\theta_{13}}-4\sin^4{\theta_{23}}\cos^4{\theta_{13}}+1.6\sin^2{\theta_{23}}\cos^4{\theta_{13}})\sin^2{x}.
\ee
The Super-K upward muon analysis requires $\s_{atm}>0.8$.  Then we have the second constraint
\be
0.8<2.4\sin^2{\theta_{23}}\cos^2{\theta_{13}}-4\sin^4{\theta_{23}}\cos^4{\theta_{13}}+1.6\sin^2{\theta_{23}}\cos^4{\theta_{13}}<1.0.
\label{nm}
\ee

Shown in figure~2 is the allowed regions given by eqns.~\ref{ne}, \ref{nm} for $\epsilon=0.05,0.1$ and $0.2$.  The overlap of the constraints gives the combined allowed region.  We can see that while $\s_{23}$ is constrained to be quite large, there is much less constraint on $\s_{13}$ consistent with the data. In particular, for $\s_{23}=1$, $\s_{13}$ could be as large as 0.2 for $\epsilon=0.05$, or as large as 0.8 for $\epsilon=0.2$.  Such solutions can occur when $\nue$ depletion is canceled by replenishment through $\nm$ oscillation to within an amount $\epsilon$.  We conclude that $\s_{13}$ could be much larger than concluded from the standard analysis. 

\FIGURE[t]{\centerline{ \epsfxsize=5in \epsfbox{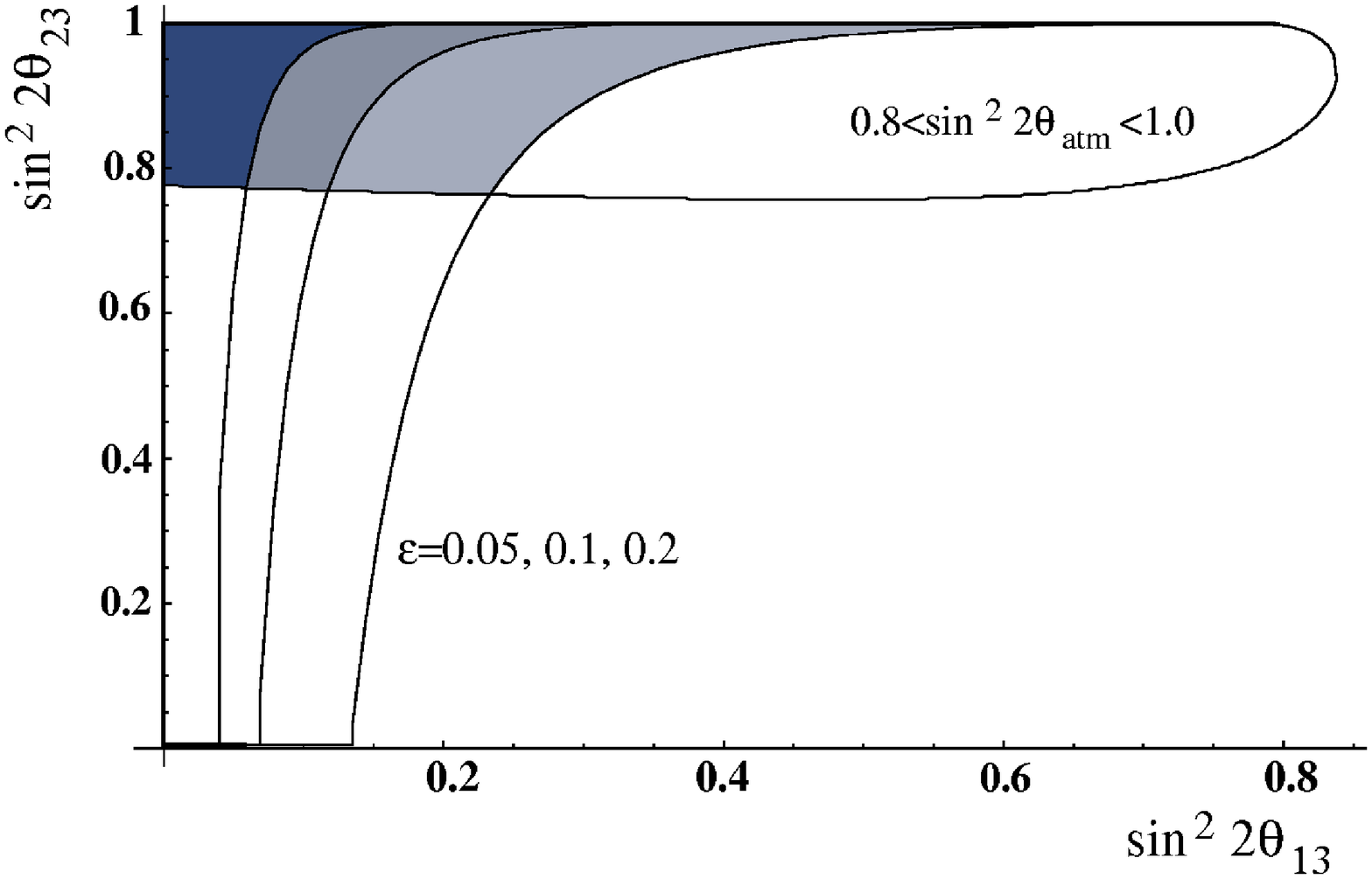} }
\caption{The curves marked $\epsilon=0.05,0.1,0.2$ are the constraints from $\nu_e$ disappearance from Super-K given by eqn.~(3.6) for 5\%, 10\% and 20\% uncertainty in the neutrino flux.  The upper curve is the constraint for $\s_{atm}$ from eqn.~(3.9).  The shaded regions are the overlap of these constraints which gives the allowed values of $\s_{23}$ and $\s_{13}$ from Super-K.  Dark shaded is for $\epsilon=0.05$, medium shaded for $\epsilon=0.1$, and light shaded for $\epsilon=0.2$.}
}

A large $\s_{13}$ would also indicate that KamLAND could observe very significant $\nue$ oscillation to the Super-K mass splitting.  This possibility is excluded in the standard three neutrino model through the limits of the CHOOZ experiment, an experiment mostly in air, $\s_{13}< 0.1$.  As we have shown here, however, the new best limits from Super-K, KamLAND, and Palo Verde cannot exclude this possibility in HDM, and in this case a three neutrino analysis of the KamLAND data would be necessary. 

To this point we have been assuming that Super-K and KamLAND are observing separate mass splittings, as the Super-K best fit point lies around $3 \times 10^{-3} \me^2$ and the KamLAND best fit around $5\times 10^{-5}\me^2$.  But we also note that the new limits from Super-K upward muon set on $\nm$ disappearance and the KamLAND data nearly overlap in $\m$.  One could ask then whether the two signals could be accommodated with a single mass splitting in HDM.  Now KamLAND reports a signal with $\s_{ex}>0.4$.  Such a large mixing for $\nue$ disappearance would have to consistent with the Super-K lack of observation of $\nue$ disappearance.  As we have shown, $\s_{13}=\s_{ex}>0.4$ is allowed for $\epsilon \geq 0.1$, which may very well be consistent with the Super-K result.  We conclude that we cannot rule out the possibility that KamLAND and Super-K are observing the same mass splitting.  But again we note that although it appears that the fits of both experiments can be accommodated with a single splitting, it is only accommodated through less preferred values of the mass splitting, as the Super-K upward muon best fit lies around $3\times 10^{-3} \me^2$ and the KamLAND best fit around $5\times 10^{-5}\me^2$.



To narrow these allowed regions, a reanalysis of the full Super-K data set would need to be carried out allowing six free parameters on $\nm$ and $\nue$ disappearance, as listed in eqn.~(3.1).  Once this analysis has been carried out, one can determine whether the combination of these parameters with the Palo Verde limits constrain $\theta_{13}$ such that a three neutrino analysis of KamLAND is necessary.  In addition, in the near future we will have additional signals when the MINOS and OPERA experiments measure $\m_{HDM},\ \s_{23}^{HDM}$ through direct detection of the {\em appearance} of $\nm\ra\nt$.  MINOS will also be able to probe $\s_{13}^{HDM}$ through $\nm\ra\nue$ down to mixing angles $\s_{13}^{HDM}\sim0.1$.  We have shown here that, with these new matter effects, the value of $\s_{13}^{HDM}$ could potentially be much larger than is now thought.  Should MINOS find a value of $\s_{13}$ in HDM which is larger than allowed by CHOOZ experiment, we have a smoking gun for new matter effects in three neutrino models.

\section{Models Consistent with LSND}

Next we consider how these new matter effects improve the fit of 3+1, 2+2 and three neutrino models to the data with LSND.  Shown in fig.~3a is the LSND allowed region for $\nm\ra\nue$ oscillations, along with the best constraints for $\nue\ra\nu_x$ and $\nm\ra\nu_x$ oscillations from Bugey and CDHS.  The modified constraints in HDM are shown in fig.~3b.  The Palo Verde experiment, not Bugey, constrains $\s_{e}$ in HDM.  The limits from CDHS on $\nm$ disappearance in HDM are unknown.\footnote{My thanks to Joe Rothberg and his collaborators at CERN for looking into beam path information.  The lapse of time between the experiment and the present, however, has not allowed them to track this information down.}  In addition, the KARMEN experiment, which is $\sim50\%$ air no longer constrains LSND.   We consider the implications of these new limits in HDM on 3+1 and 2+2 models with LSND, which are disfavored with standard interactions.  We also consider how a three neutrino model with LSND becomes workable with non-standard interactions in HDM.


\FIGURE[t]{
\centerline{a)\ \ \epsfxsize=3.5in \epsfbox{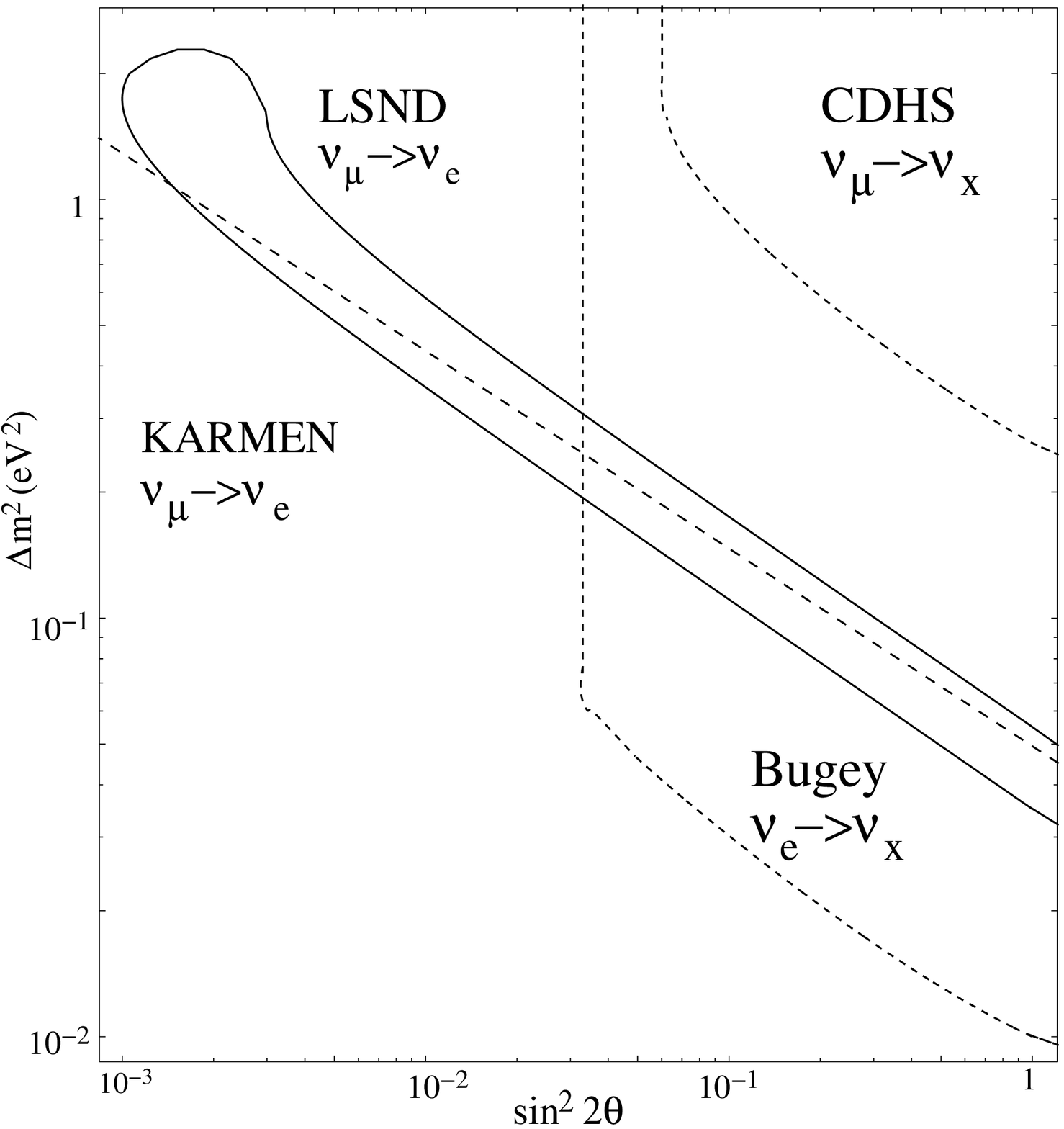} \hskip 0.25in b)\ \epsfxsize=3.5in \epsfbox{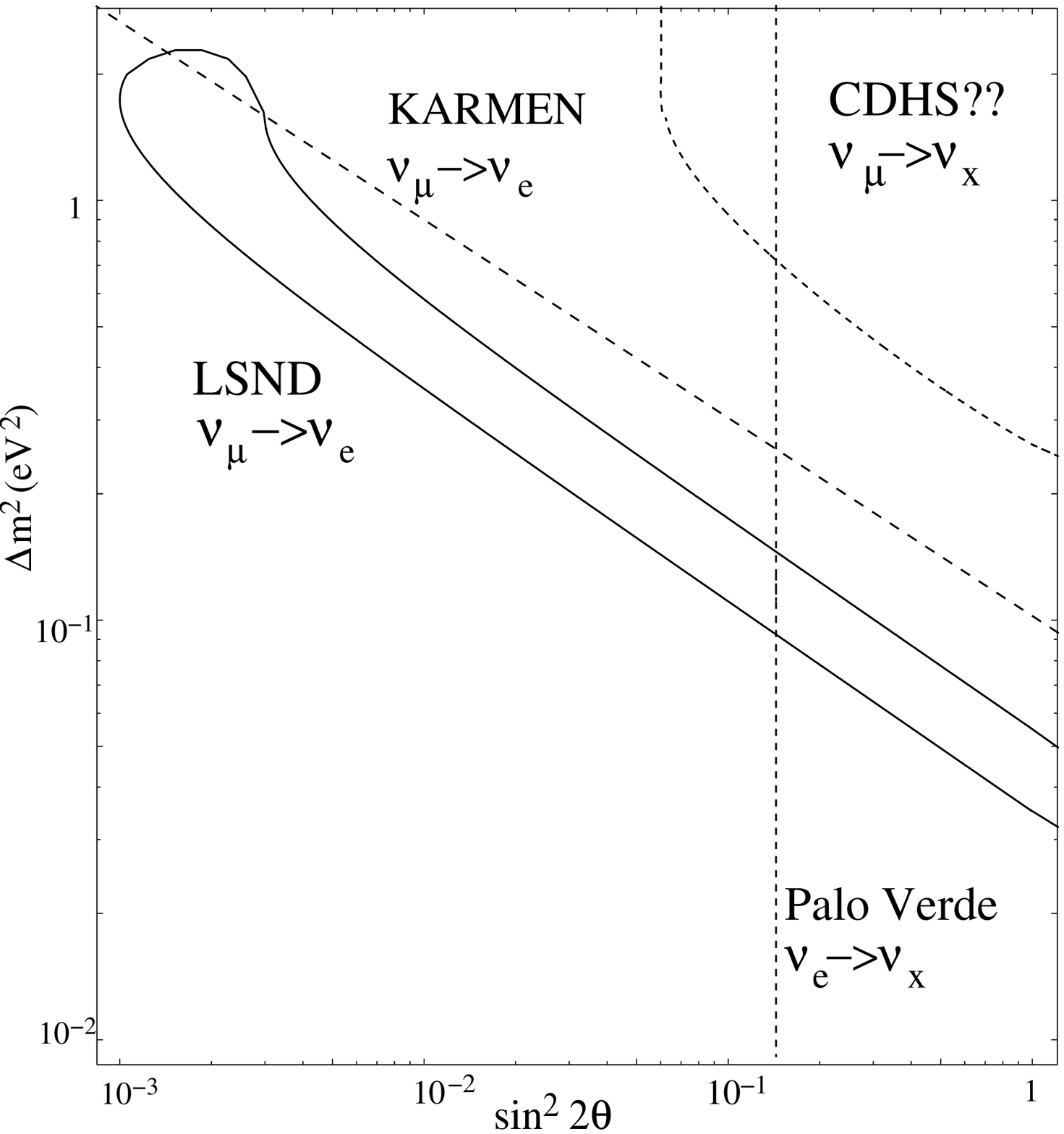} }
\caption{LSND allowed region along with constraints on $\nue$ and $\nm$ oscillations from null searches a) Bugey and CDHS with standard interactions and b) Palo Verde and CDHS with non-standard in-medium interactions.  As the fraction of the CDHS beamline in HDM is unknown, the limits for non-standard interactions are at most as restrictive as those shown.  The curves are parameterizations of the published allowed regions; in addition irregularities in the Bugey curve for large $\m$ have been suppressed, as they are irrelevant for the analysis conducted here. Dashed lines indicate limits from null searches, and solid lines allowed regions from positive signals.}
}

\subsection{3+1 models}

These medium effects improve the fit of 3+1 models to the KamLAND, Super-K, Bugey, Palo Verde, and LSND results. 

In typical 3+1 models, we have three splittings, $O(10^{-5}\me^2)$, $O(10^{-3}\me^2)$, and the larger LSND splitting $O(0.1-1 \me^2)$. With only standard interactions, 3+1 models are disfavored \cite{Maltoni:2001bc}, and a 3+2 scenario \cite{Sorel:2003hf} improves the fit somewhat, but only by contriving to set the masses of the sterile neutrino in the region where the constraints from CDHS and Bugey are weakest.  

The reason that these models are disfavored is that the combination of the Bugey and CDHS limits on $\nue$ and $\nm$ result in a $\mu-e$ mixing angle which is too small at $90\%$ C.L. to accommodate the LSND result.  To quantify this observation, we write the $\nue$ and $\nm$ disappearance angles measured by Bugey and CDHS in terms of the fraction of $\nue$ or $\nm$ in the LSND mass splitting, $U_{e4}$ and $U_{\mu4}$:
\be
\s_{e}=4(1-\mid U_{e4}\mid^2)\mid U_{e4}\mid^2
\ee
\be
\s_{\mu}=4(1-\mid U_{\mu4}\mid^2)\mid U_{\mu4}\mid^2.
\ee 
Likewise, the LSND $\mu-e$ mixing angle is written in terms of these same quantities
\be
\s_{\mu e}=4\mid U_{e4}\mid^2\mid U_{\mu4}\mid^2=(1-\cos 2\theta_{\mu})(1-\cos 2\theta_e).
\ee
When the limits from Bugey and CDHS on $\s_e$ and $\s_\mu$, as shown in fig.~3b, are substituted into these expressions, we find that the resulting allowed mixing $\s_{\mu e}$ is too small at 90\% C.L. to accommodate the LSND result. There is only a small region of LSND parameter space, $0.2 < \m < 0.3$, which, although excluded at $90\%$ C.L. is allowed at $99\%$ C.L.  Hence these fits are only marginally allowed within a small region of parameter space. 

We immediately see how the situation changes with these new matter effects.  Bugey, an experiment mostly in air, can no longer constrain LSND.  Instead, the constraint on $\nu_e$ disappearance in HDM comes from Palo Verde $\sin^2 2\theta_e <0.17$, which corresponds to the region of LSND parameter space $\Delta m^2>0.1\mbox{eV}^2$. These limits are considerably less restrictive than $\sin^2 2\theta_e <0.04$ from Bugey, and the result is that the combined limits of Palo Verde and CDHS do not disfavor the LSND result.  

The Palo Verde limit implies that if MiniBooNE sees $e-\mu$ oscillations, it should see them with $\Delta m^2>0.1\mbox{eV}^2$.  Furthermore, if oscillations are detected at MiniBooNE in the region $0.1\mbox{eV}^2<\Delta m^2<0.25\mbox{eV}^2$ (excluded by Bugey in a conventional analysis), we have strong indication for non-standard matter interactions for neutrinos. 

An additional prediction of this model is that a reactor experiment, with similar sensitivity as Bugey, but with neutrino pathlengths instead in HDM, would detect $\nu_e$ disappearance, provided the CDHS limits are rigorous in rock.  The reason is that the combination of CDHS and Bugey limits with standard interactions disfavor at 90\% C.L. mixings large enough to accommodate the LSND result.  We would expect a signal in the approximate range $0.04<\sin^2 2\theta < 0.17$, $0.1 \me^2<\Delta m^2 < 1 \me^2$.   The caveat here is that the CDHS limits be applicable in HDM; or, equivalently, that some other experiment (e.g. Super-K) limit $\nm$ disappearance in HDM, for the mass splittings of interest for LSND, at least as much as CDHS for a given $\m$.  Currently the material in the neutrino path of CDHS is unknown.  If the experiment had been partially in air, such that ${\m}^{CDHS}_{min}/r \gtrsim 0.3$, where r is the fraction of the neutrino pathlength in HDM, and no other experiment places as tight of limits on $\nm$ disappearance as the standard CDHS limits, we would have no prediction for a signal from such a reactor experiment.  

\subsection{2+2 models}

Analysis of the SNO, KamLAND, and Super-K data strongly disfavors 2+2 oscillation schemes in the standard interaction picture  \cite{Maltoni:2001bc,Habig:2001ej,Ahmad:2002ka,Bahcall:2002zh,deHolanda:2002ma} because 2+2 models with LSND require large mixing with sterile states, which are disfavored by the SNO, KamLAND and Super-K atmospheric data.

To determine the fit of 2+2 models to the data in HDM, a new analysis would need to be carried out.  The existing analyses distinguish between active-active versus active-sterile oscillations through the MSW effect, which couples only to active neutrinos. These analyses, however, are only valid if the MSW effect is the dominant matter effect, and hence do not constrain active-sterile mixing in the case that the new non-standard interactions considered here dominate.  A new analysis with the new interactions, however, would be highly model dependent, as the form of the potential due to the new interactions must be known.  As we have been carrying out only a model independent analysis for such new interactions, we will not perform such an analysis here.  They could, however, easily be carried out once the form of the potential is known. 

\subsection{Three neutrino models}  Traditionally, the LSND result has been interpreted to necessitate a fourth, sterile, neutrino because consistency with the Bugey experiment requires $\m_{LSND}\gtrsim 0.3 \me$, but $\m_{atm}\sim 3\times 10^{-3} \me$, and  $\m_{solar} \sim 7\times 10^{-5}$.  However, as table~\ref{table1} indicates, the limits from Super-K in HDM are modified so that $\m_{atm}\lesssim 6 \times 10^{-3}\me$ and the limits from LSND so that $\m_{LSND}\gtrsim 0.1 \me$.  This Super-K upward muon data set, however, though containing mostly HDM pathlengths, also contains a horizon bin, whose pathlength is mostly through air.  Without a reanalysis of the data where bins containing pathlengths mostly through air are removed, we cannot be sure that the allowed region is an accurate reflection of the parameters in HDM.  Furthermore, the allowed LSND and Super-K regions are approximately an order of magnitude apart.  Given such sources for systematic error in the analysis and the proximity of the allowed regions, it is interesting to consider models in which both the LSND and Atmospheric results in rock are accommodated with the same mass splitting.

In \cite{Kaplan:2004dq}, a model was considered in which the atmospheric and LSND mass splittings were accommodated by a splitting of $O(0.1 \me^2)$ in rock, while in air (where the LSND result is not applicable) the splitting is $O(10^{-3} \me^2)$, and the KamLAND result by a splitting of $O(10^{-4}\me^2)$.  In this model, the upper state is described by 
\be
|\nu_3\rangle = \frac{1}{\sqrt{2}}(|\nm \rangle + |\nt \rangle) +\epsilon|\nue \rangle),
\ee
to describe the large mixing atmospheric results, with a small admixture of $\nu_e$ for the LSND result, and the lower two states by 
\be
|\nu_1\rangle = \frac{1}{\sqrt{2}}|\nue \rangle + \frac{1}{2}|\nm \rangle -\frac{1}{2}|\nt \rangle
\ee
and
\be
|\nu_2\rangle = -\frac{1}{\sqrt{2}}|\nue \rangle + \frac{1}{2}|\nm \rangle -\frac{1}{2}|\nt \rangle.
\ee

We discuss here on how such a model would be directly testable.  First, in the very near future K2K would be able to rule out such a model if it detected high statistics $\nm$ disappearance with a mass splitting in the range $\m<0.1 \me^2$. 

The MINOS and OPERA experiments will provide another means to measure this same mass splitting in rock in the near future.  Additionally, MINOS will have the capability of measuring $\nm-\nue$ oscillation down to mixings of order, $\s_{\mu e}=4\times 10^{-2}$ which are interesting for LSND mass splittings $\m<0.4\me^2$. 

Like the 3+1 model, this three neutrino model also predicts that a reactor experiment with similar sensitivity as Bugey, but with pathlengths instead in HDM, would detect $\nue$ disappearance in the same parameter range of interest.

\section{Summary}

\begin{table}
\hspace{-1.1in}
\begin{tabular}{l|l|c|c|c|c|c|c} \hline \hline
Model & Parameter & \multicolumn{2}{c|}{SI range} & Experiment &\multicolumn{2}{c|}{HDM range} & Experiment \\ \hline
3 neutrino & $\s_{13}$ & 0 & 0.1 &CHOOZ& 0 & 0.5 & Super-K, KamLAND  \\
&&&&&&&K2K,MINOS,Reactor Expt.\\ \cline{2-8}
&$\s_{12}$ & 0.76&0.88  &SNO&0.4&1&KamLAND \\ \cline{2-8}
&$\s_{23}$ & 0.92& 1&Super-K&0.8&1&Super-K,K2K  \\
&&&&&&&MINOS,OPERA \\ \cline{2-8}
&$\m_{12}$&$6.5\ten{-5}$&$8.2\ten{-5}$&SNO&$10^{-3}$&$10^{-5}$&KamLAND \\ \cline{2-8}
&$\m_{23}$&$1.9\ten{-3}$&$3.0\ten{-3}$&Super-K&$1\ten{-3}$&$6\ten{-3}$&Super-K,K2K \\ 
&&&&&&&MINOS,OPERA \\ \hline
3+1 neutrino & $\su_{\mu e}$& \multicolumn{2}{c|}{disfavored} &CDHS,Bugey&$10^{-3}$ &$10^{-1}$&MiniBooNE,Reactor expt.\\ \hline
2+2 neutrino & $\su_{\mu s}$& \multicolumn{2}{c|}{disfavored} &Super-K& 0&1&MINOS,OPERA \\ \cline{2-8}
& $\su_{e s}$& \multicolumn{2}{c|}{disfavored} &SNO&0 &1&no experiment \\ \hline \hline

\end{tabular}
\caption{Parameters of interest for each type of model.  Standard interactions (SI) upper and lower limits (90\% C.L.) and the experiment which gives that limit (columns three and four).  Present HDM limits and experiments which are of interest for measuring these parameters in the near future (columns five and six).}
\label{table2}
\hspace{1.1in}
\end{table}

We have examined the effects on neutrino oscillations of a very light scalar mediating a force between neutrinos and ordinary matter, as proposed in [5,6].  As these new matter effects could be much larger than the MSW effect, we have considered how these new interactions would significantly modify the constraints on the neutrino parameters in any High Density Medium (HDM).  I summarize in table~\ref{table2} for each parameter the allowed range and experimental signals with standard interactions versus these new in-medium non-standard interactions.  In particular, indications for these new matter effects from already planned and running experiments include:
\begin{itemize} 
\item $\s_{13}^{HDM}$ with a value larger in HDM than the limit given by the CHOOZ experiment;
\item $\m_{12},\ \s_{12}$ detected by KamLAND in a range not allowed by the SNO results;
\item $\m_{23},\ \s_{23}$ detected by K2K, OPERA, and MINOS with parameters somewhat modified from the Super-K result;
\item  Detection of $\nue$ disappearance in HDM in the region excluded by Bugey, should MiniBooNE confirm the LSND signal.
\end{itemize}
Any of these signals would be a smoking gun for the new matter effects considered here.

These upcoming and currently running experiments promise to determine the parameters quite accurately in rock.  We have, however, no signal for the parameters in air, as well as no information on the form of the density dependence.  Without a theory for the density dependence of such a force, we are unable to constrain the parameters from solar neutrinos, and the constraints from short baseline and atmospheric neutrino experiments could be significantly different if the neutrino parameters are sensitively dependent to small variations in density.  

Experiments are needed to determine if and how sensitively the neutrino parameters vary from such a new force.  In the past it has been assumed that new matter effects were arising from physics above the standard model energy scale, and for typical earth densities were unimportant.  It is suggested here that a new attitude be taken in which large non-standard in-medium dependences are searched for in terrestrial neutrino experiments.  We may find that terrestrial neutrino oscillations provide the tool to directly explore a wide range of physics beyond the standard model. 

\acknowledgments
I am grateful to David Kaplan, Ann Nelson and Neal Weiner for many
helpful comments along the way, and for proposing the problem in the
first place. Much thanks also to Jeff Wilkes, Richard Gran and Kiyoshi
Shiraishi for useful discussions on Super-K, and to Rob Fardon for reading various drafts and making useful suggestions.  This work was supported by the US Department of Energy grant DE-FG03-00ER41132.

\bibliography{nuexpmt}
\bibliographystyle{jhep}
\end{document}